\documentclass[12pt,a4paper]{article}
\usepackage{amsfonts,amsmath,amssymb,amscd,epsfig}

\begin{document}

\title{"Particle-like"~singular solutions in  Einstein-Maxwell theory  and in algebraic dynamics}
\author{V.V.Kassandrov$^1$, V.N.Trishin$^2$ \\
$^1$ Russian People's Friendship University \\ E-mail: vkassan@mx.pfu.edu.ru 
\\ $^2$ Moscow State University}
\date{~}
\maketitle
\begin{abstract}
Foundations of {\it algebrodynamics} based on earlier proposed equations of {\it
biquaternionic holomorphy} are briefly expounded. Free Maxwell and Yang-Mills Eqs. are
satisfied identically on the solutions of primary system which is also related to the Eqs. of
{\it shear-free null congruences} (SFC), and through them - to the Einstein-Maxwell
electrovacuum system. {\it Kerr theorem} for SFC reduces the basic system to one algebraic
equation, so that with each solution of the latter some (singular) solution of vacuum Eqs.
may be associated. We present some exact solutions of basic algebraic and of related field
Eqs. with {\it compact} structure of singularities of electromagnetic field, in particular
having the form of {\it figure "8"~curve}. Fundamental solution to primary system  is
analogous to the metric and fields of the {\it Kerr-Newman} solution. In addition, in the
framework of algebraic dynamics the value of {\it electric charge} for this solution is
strictly fixed in magnitude and may be set equal to the {\it elementary charge}.
\end{abstract}

\section{\hspace{-4mm}.\hspace{2mm}Introduction}

As the main goal of the {\it algebrodynamical} approach \cite{m}, we regard the derivation of
equations of physical fields (or of fundamental physical laws, in a broader sense) from a
unique primary
 {\it Principle of purely abstract nature}, based only on the intrinsic properties of exclusive
mathematical structures (groups, algebras, mappings etc.). Basic realization of this concept
developed in \cite{gc} makes use of only the {\it differentiability conditions} of functions
of {\it biquaternionic variable}, i.e. of the Cauchy-Riemann (CR)  equations
generalized to the case of noncommutative (associative) algebras.

In the framework of the version of analysis proposed in \cite{m} the noncommutativity of
starting algebra naturally results in the nonlinearity of generalized CR equations (GCRE),
justifying the use of the latters as the dynamical equations of {\it interacting} fields.

Infact, for the algebra of biquaternions {$\mathbb B$} (isomorphic to the full $2\times2$
complex matrix algebra) the GCRE appear to be Lorentz invariant and carry natural 2-spinor
and gauge structures. In the most examined case for which a simple geometric and physical
interpretation is obvious, the conditions of {\it $\mathbb B$-differentiability} reduce to
the invariant system \cite{gc} $$ \eta(x+dx)-\eta(x) \equiv d\eta=\Phi(x)*dX*\eta(x) \eqno(1)
$$ for $\mathbb C$-valued 2-spinor $\eta(x)$ and gauge $\Phi(x)$ fields represented by
$2\times1$ and $2\times2$ $\mathbb C$-matrices respectively. $(*)$ in (1) denotes usual
matrix multiplication (equivalent to that in $\mathbb B$), and $dX$ represents a $2\times2$
Hermitian matrix of the increments of coordinates.

As a consequence of (1) each component of the 2-spinor $\eta _A(x) (A,B=1,2)$ satisfy the
4-eikonal equation \cite{vest}. On the other hand, the {\it compatibility conditions} $d\wedge
d\eta=0$ impose dynamical restrictions on the gauge field $\Phi(x)$. Namely, the $2\times 2$
matrix $\mathbb C$-valued {\it connection 1-form}
$$\Gamma(x)=\Phi(x)*dX=\Gamma^0(x)+\Gamma^a(x)\sigma_a \eqno(2) $$ ($\sigma_a,~ a=1,2,3$
being the Pauli matrices) by virtue of compatibility conditions {\bf should be self-dual}.
Consequently, {\bf free Maxwell and Yang-Mills equations are satistied} identically on the
solutions of (1), for the scalar $\Gamma^0(x)$ and the vector $\Gamma^a(x)$ parts of the
connection (2) respectively \cite{gc}. Thus, the GCRE system (1) exhibit wonderful relations
to several fundamental physical equations being for the latters in some sense generating.

In the approach regarded, an important role is performed by {\it singular sets} where the
strengths of Maxwell and YM fields turn to infinity. Such sets were found to have diverse
dimensions and topology. Solutions with {\it compact} structure of singular set may be then
considered as {\it particle-like}, the evolution of such singularities-particles being then
governed by the system (1) itself.
\medskip

In this paper, we expound the process of reduction of system (1) to the equations of {\it
shear-free geodesic null congruences} and, by Kerr theorem, to the solution of {\it
algebraic} equation  \cite{kazan97}. Making use of this, we present an explicit form of the
solutions to free Maxwell equations, analyze the structure and evolution of their
singular set and discuss the relation of these electromagnetic fields to the Kerr-Shild metrics
and, by this, to the solutions of electrovacuum Einstein-Maxwell equations. In conclusion, we
consider general status and physical interpretation of "particle-like"~singular
solutions in the framework of electrogravidynamics and of unified algebraic
field theory proposed.

\section{\hspace{-4mm}.\hspace{2mm}Reduction of GCRE system and the solutions to
free Maxwell equations.}

Through the elimination of the gauge field $\Phi(x)$ the system of GCRE (1) may be written in
a 2-spinor form \cite{kazan98} $$ \eta_{A^{'}}\nabla^{AA^{'}}\eta^{B^{'}} =0     \eqno(3) $$
In the gauge $\eta^{A^{'}}(x)= (1,G(x))$ the system (3) reduces to two equations for one
unknown function $G(x)$ $$ \partial_{\bar w}G=G\partial_uG, \qquad \partial_vG=G\partial_w G,
\eqno(4)$$ $u,v= t \pm z, \quad w,\bar w=x \pm i y$ being the {\it spinor coordinates}.
Note that as a consequence of (4) $G(x)$ satisfy identically both the 4-eikonal and
d'Alembert wave equations \cite{kazan98}. Assuming the Eqs. (4) are solved, the components of
4-potential matrix $\Phi(x)=A_0+A_a(x)\sigma_a$ may be expressed through the function $G(x)$
as $$A_w=\partial_uG,~~ A_v=\partial_wG, ~~A_u=A_{\bar w}=0   \eqno(5)$$ and satisfy free
Maxwell equations.

Wonderfully, Eqs.(4) are completely identical to the equations of {\it shear-free null geodesic
congruences} in the gauge for the spinor $\eta(x)$ regarded \cite{penrose}. In accord with
Kerr's theorem \cite{penrose}, we then obtain the general solution of Eqs.(4) in an implicit algebraic form
$$F(G, \bar wG+u, vG+w)=0, \eqno(6)$$ $F(G,\tau_1,\tau_2)$ being an arbitrary golomorphic
function of three complex variables including two {\it twistor} (projective) components
$$\tau_A =X_{AA^{'}}\eta^{A^{'}}, \quad\tau_1=\bar w G+u,~ \tau_2=vG+w \eqno(7) $$ Thus, a
lot of solutions to free Maxwell equations may be obtained through simply examining of the
algebraic Eq.(6). Singularities of related field strengths may be then found from the {\it
caustic} condition $$\frac{dF}{dG}\equiv \partial_G F+\bar w \partial_{\tau_1} F
+v\partial_{\tau_2} F=0 \eqno(8)$$ Eliminating the only unknown function $G(x)$ from two
algebraic Eqs.(6),(8), one easily comes to the {\it equation of singular set} which
determines the shape and evolution of singularities, without even taking care of explicit
solving the Eq.(6) itself. The example of such procedure was presented in \cite{KW}.

\section{\hspace{-4mm}.\hspace{2mm}Stationary solutions.}

From the structure of twistor components (7) it's evident that stationary solutions to Eq.(6)
are exhausted by the functions $F(G,\lambda)$, where $\lambda = G\tau_1 - \tau_2$ doesn't
contain the time variable $t=\frac{1}{2} (u+v)$ at all. In accord with the results of Kerr
and Wilson \cite{KW}, for stationary solutions with {\bf compact} structure of singular set
the function $F$ {\bf should be at most quadratic in $G$}, i.e. should have the form
$F=(G\tau_1-\tau_2)+ a_0G^2+a_1G+a_2, \quad a_i\in \mathbb C$. Linear dependence on $G$
immediately leads to the trivial solution with zero fields. Using 3-translations and
3-rotations, the above form may be reduced to $F=G\tau_1-\tau_2-2aG$ for which we obtain from
quadratic Eq.(6) $$G(x)=\frac{x+iy}{z-a\pm\sqrt{(z-a)^2+x^2+y^2}} \eqno(9)$$ For a real
valued $a$ from (9) and the expression for potentials (5) we obtain  the Coulomb
electric field with a point singularity and a {\bf fixed value of the electric charge} as a
consequence of nonlinear primary system of GCRE (1). Imaginary values of $a$ correspond to
the {\it ring singularity} with radius $r=|a|$ and multipole structure of EM fields with
Coulomb first main term \cite{kazan97}. Decomposition and separation of real parts of EM
fields at a
 distance $r\gg |a|$ gives $$E_r
\simeq \frac{e}{r^2} (1 -\frac{3 a^2}{2 r^2}(3\cos^2{\theta}-1)), E_\theta \simeq -\frac{e
a^2}{r^4} 3\cos{\theta}\sin{\theta}, $$ $$H_r \simeq \frac{2 e a}{r^3}\cos{\theta},\ H_\theta
\simeq \frac{e a}{ r^3} \sin{\theta}. \eqno(10)$$ Contrary to an ambiguous value of ring's
radius $a$, the dimensionless electric charge is strictly fixed (up to a sign) by field
equations, so that in absolute units it may be identified with {\it elementary charge} $e$.
Note that the solution (9) (for the case $a=0$) and the property of charge quantization for the GCRE system (1)
have been obtained in a direct way in \cite{m,gc}. (Recently \cite{ranada, zhour} there
were some interesting attempts to explain electric charge quantization by
topological reasons instead of dynamical considerations used here).

On the other hand, solution (9) is extremely important in the framework of GTR. Indeed, the
expression $l_\mu = \eta^+ \sigma_\mu \eta$, where $\eta^T = (1,~G(x))$ is the 2-spinor
related to the function  $G(x)$, defines the {\it principal null congruences}~ $l_\mu$ of a
Riemannian space-time endowed with a Kerr-Shild metric $$g_{\mu\nu} = \eta_{\mu\nu} +
H(x)~l_\mu l_\nu, \eqno(11)$$ where $\eta_{\mu\nu}$ represents the metric of auxiliary
Minkowski space-time. The scalar factor $H(x)$ should be then determined by the Einstein
vacuum or electrovacuum equations and for fundamental stationary solution (9) leads to the Kerr
or Kerr-Newman metrics respectively (consequently, to the Schwarzschild or Reissner-Nordstr{\"o}m
metrics for the case $a$=0). In our approach, it is of great importance that {\bf singularities of
curvature of metric (10) are fixed \cite{KW, B80} by the same condition (8) as for
electromagnetic field} and define infact one unique particle-like object. Another wonderful
fact is that for the Kerr-Newman solution of Einstein-Maxwell equations electromagnetic fields are
just those defined by the GCRE system (apart from the property of quantization of charge for
the latters!) and may be asymptotically presented by the Eq.(10). Moreover, these fields {\bf
obey Maxwell equations both in flat space and in Riemannian space with Kerr-Shild metric (11)!}
Such a remarkable {\it property of stability} of electromagnetic fields under Kerr-Shild
deformations of space-time geometry noticed in \cite{abstr} will be discussed elsewhere.

Making use of correspondence between the fundamental solution (9) to the
GCRE system (1) and the Kerr-Newman solution of Einstein-Maxwell system, one
is able to endow the solution (9) with a complete set of quantum numbers
(including mass and spin). Then the {\it gyromagnetic ratio} would automatically
correspond to that for Dirac particle, while the charge would be fixed in
magnitude. Unfortunately, no natural reasons to ensure the quantization of
mass could be seen nowadays. We'll continue the discussion below.

\subsection{\hspace{-5mm}.\hspace{2mm}Nonstationary solutions.}

Let us consider the general quadratic form of the function $F(G,\tau_1,\tau_2)$. When the
terms bilinear in $\tau_1$, $\tau_2$ are absent, such functions (under the
restriction on singular set to be compact) correspond to the boosted or rotated Kerr solution
\cite{bur3}. On the other hand, the function $F=\tau_1\tau_2-b^2G$ has been considered in
detail in \cite{kazan98}. The explicit expression for $G$ in this case is $$G = \frac{-2u
w}{\sigma^2 + \rho^2 + b^2 \pm \sqrt{\Delta}},~~~~\Delta \equiv (\sigma^2 + \rho^2 + b^2)^2 -
4\sigma^2 \rho^2, \eqno(12)$$ where $\sigma^2=uv=t^2-z^2,~~\rho^2=w\bar w=x^2+y^2$.  EM
fields correspondent to (12) are $$E_\rho =\mp \frac{8b^2\rho z}{\Delta^{3/2}}, ~~E_z = \pm
\frac{4b^2} {\Delta^{3/2}}(t^2-z^2+\rho^2+b^2), ~~H_{\varphi} = \mp \frac{8b^2\rho t}
{\Delta^{3/2}}. \eqno(13)$$ For real $b$ the fields (13) are identical to the well-known {\it
Born solution} for two point-like charged "particles" performing uniformly accelerated
counter-motion. The value of electric charge for each particle does not depend on $b$ being
{\bf fixed and equal to the charge of fundamental solution (9)}. For the case of imaginary $b$ one
has the singularity of rather exotic toroidal structure, defined by the equation
$z^2+(\rho\pm b)^2=t^2$ (see \cite{kazan98} for details). In general case of complex-valued
$b$ singular set manifests itself as the two rings of fixed radii performing again the
oncoming hyperbolic motion along $z$-axis.

It may be proved that (up to the transformations of Poincare group) the {\it axisymmetric}
solutions to the Eq.(6) (and to the GCRE system (1) respectively!) generated by
quadratic function $F$ {\bf are exhausted by the Kerr-like solution (9) and the nonstationary
bisingular solution (12)} together with (toroidal or double ring-like) modifications of the
latter.

It seems, however, that the solutions to GCRE with compact singularity and non-axial
symmetries may be of interest too. Here we present an example of such solution which may be
obtained from the generating function $F=\tau_1\tau_2-a^2G^2$. Resolving the equation $F=0$,
one comes to the following expression $$G=\frac{2uw}{\pm\sqrt{\Delta}+uv+w\bar w},
\eqno(14)$$ $\Delta$ where  $\Delta = (t^2-x^2-y^2-z^2)^2-4a^2(t+z)(x+iy)$. The singular set
for this solution is defined by the condition $\Delta=0$  and for $t=0$ has the form of flat
{\it figure "8"~ curve} (Fig.a)~). The time evolution of this singularity is illustrated by
Fig.b).

\begin{figure}[ht]
a)~t=0~~\epsfig{file=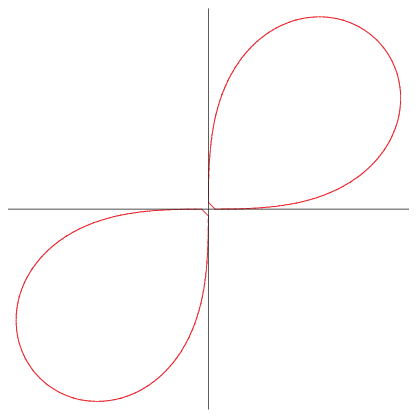, width=2in, height=2in, angle=0}~~~~~b)~t=1~~
\epsfig{file=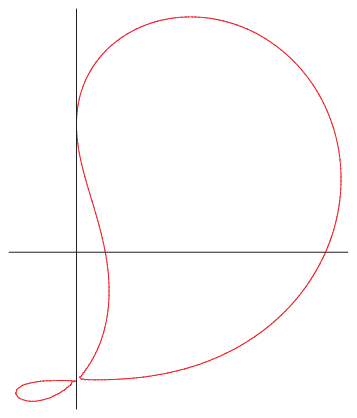, width=2in, height=2in, angle=0}
\label{pic1}
\end{figure}

EM fields related to the solution (14) and being represented by the complex combination $\vec
{\mathcal{E}} = \vec E - i\vec H$ are:

$${\mathcal{E}}_+\equiv {\mathcal{E}}_1+i{\mathcal{E}}_2=-\frac{2a^2w^2}{\Delta^{3/2}},
~~~{\mathcal{E}}_-\equiv {\mathcal{E}}_1-i{\mathcal{E}}_2=\frac{2a^2u^2}{\Delta^{3/2}},
~~~{\mathcal{E}}_3=-\frac{2a^2uw}{\Delta^{3/2}}. \eqno(15)$$

For  each finite moment of time they decrease rapidly (as $r^{-4}$) with the distance from
the centre of singularity. The fields are neutral (with total charge being equal to zero) and null
($\vec E^2 - \vec H^2 = 0,~~\vec E * \vec H = 0$).

In conclusion, we present a peculiar solution with noncompact singularity
which serves as the {\it analogue of electromagnetic wave} in GRCE dynamics. For
the solutions of wave-like type the generating function $F$ should depend only
on one twistor component, say, $\tau_1$. Then, for the equation $F(G,\bar w G+u)=0$
the initial distribution of $G(u)$ may be arbitrary fixed at $\bar w=0$, i.e.
at the $Z$-axis. Choosing for the latter the {\it monochromatic} dependence and
resolving the equation $G - A \exp{i \Omega(\bar w G + u)} = 0$, where
the parameters $A,\Omega$ are  assumed to be real and positive, we find

$$G = iW(-iA \Omega \bar w \exp{ i \Omega u}) / \Omega \bar w, \eqno(16)$$
$W$ being the principal branch of the so called {\it Lambert}~ function which
is the solution of the equation $W(z) \exp{W(z)} = z$.

The structure of singular set is simply derived after that and appears to be a neutral {\it
helix} of radius $1/\Omega A e$ and of lead $2\pi/\Omega$ propagating along $Z$- direction
with the speed of light. Electromagnetic fields are mutually orphogonal and transversal while
polarization depends on the distance from the axis. In the direction perpendicular to the
axis the fields fall at large distancies as $1/r$. As before, the fields are globally defined
only up to a sign.

Shear-free null geodesic congruences $l_\mu$ and the Kerr-Shild metrics (12) may be
associated with the nonstationary solutions above-presented up to the scalar
factor $H(x)$. At present it's not clear if the latter may be choosed so that
the Einstein-Maxwell electrovacuum system would be satisfied. By this, an
interesting representation \cite{lind,bur4} for the shear-free
congruences (through consideration of null cone emanated by the source
moving along some curve in {\it complex space}) as well as the {\it condition of
stability} for electromagnetic fields under the Kerr-Schild deformations of
space-time geometry \cite{abstr} may be of great use.

\section{\hspace{-4mm}. \hspace{2mm}General status of~ "particle-like"~ singular
solutions}

Well-known are the numerous problems arising in GTR and in quantum field theory in respect to
the singularities of solutions of field equations (violation of causality
\cite{carter,penrose2}, divergences etc.). On the other hand, just the naked singularity of
Kerr-Newman solution (which appears instead of black hole solution in the case of a large
angular moment) manifests itself many remarkable properties related to that of elementary
particles. Accordingly, several attempts to construct the model of electron on the base of
Kerr-type solutions (KTS) have been undertaken \cite{carter,lopes}.

However, they all dealt with the problem of physically suitable {\it source} for
KTS to be found which is tigthly related to the well-known {\it twovaluedness} of
Kerr-like geometry and electromagnetic fields in particular. Infact, the
introduction of  source becomes admissible only after the {\it
cut} of space which restore the global uniqueness of the $\mathbb C$-valued
functions representing the fields of KTS.

Unfortunately, the surface of cut is quite ambigious: it may be either the
{\it disk} spanning the Kerr singular ring \cite{israel} or the oblate
spheroid \cite{lopes} covering the singular ring on which the Kerr-Newman
metric turns surprisingly into the
Minkowski one. Consequently, one may think of the source of KTS as of the
"rotating relativistic disk", of the "bubble of flat geometry" within the
external Kerr-Newman space-time etc. Thus, we are to conclude that {\bf there
are no grounds to speak about the "source"~ of KTS solutions at all} since the
twovaluedness is the unavoidable feature of their internal mathematical structure.

To illustrate the above statement, let us consider a simplier case of the
singular "particle-like"~ solutions to free Maxwell equations in flat space-time
presented in this
paper. Note that all of them (apart from the Coulomb and Born solutions with
point-like singularities) are of the same two-valued structure being in each
point defined up to a sign. Certainly, by no $\delta$-
function distribution of charge and current along the singular curve one can
reproduce the field distribution in the whole space.

On the other hand, such solutions are locally well defined and may be
analytically continuated from the region of regularity so that the full
structure of singular set is established {\bf in a unique way}.
One cannot in any way change either the shape and
topology of the singularity or its time evolution (the latter property
being the most important from physical point of view). Suppose we really hope to
describe the interactions and transmutations of elementary particles by means
of the solutions regarded (which are rather to be {\it multisingular} for real
physical process). Then we'll proceed in well-defined and unique predictions
in spite of partial indefinitness of EM fields and escape any divergences
at all!

Moreover, one may think of such solutions as of {\bf the only possibility to explain the
"spin 1/2"~ structure} at a purely classical level and with transparent picture of space-time
dynamics being preserved. This is still more true in the framework of the algebrodynamical
approach we develop, at least for two reasons. The first one is that in respect to the
internal structure of GCRE system (1), gauge (electromagnetic plus Yang-Mills) fields stand
there hand by hand with the 2-spinor structure so that the latter appears naturally together
with Maxwell equations. The second reason is that, apart from the right value of gyromagnetic
ratio, the value of electric charge is automatically fixed by the field equations themselves.

Of course, the stationary KTS as well as bisingular and "figure 8"~ solutions
presented here can say nothing about real dynamics of an ansamble of compact
"particle-like"~ singularities. Even the problem of interaction of two Kerr-Newman
objects is far from solution. In the framework of algebrodynamics the
overdetermined structure of GCRE impose restrictions even on the initial
distribution of the fields \cite{kazan98} so that the scattering problem should be fully
reformulated.

Historically, the solution of Maxwell equations with ring singularity have been obtained by
Appel in 1887 and revived in the works of Newman and Burinskii  \cite{newman, bur2}. General
study of singular solution to Maxwell equations have been undertaken by Bateman \cite{bateman}.
Nowadays the concept of naked singularities of KTS as the model for elementary particles is
successfully developed, say, in the works of Clement \cite{clement}.

To conclude, we argue that hostile attitude of physicists to singularities
of field equations could be quite unjustified. There exist no restrictions
of principal character for the  {\it compact multisingular} solutions to describe
the interactions of particle-like objects in a self-consistent
way. Then their transmutations could be treated as {\it perestroikas} of
singularities in terms of catastrophe theory.  This programme should be
implemented independently both in the framework of Einstein-Maxwell dynamics
and of the algebraic dynamics based in particular on the GCRE system (1).

\end{document}